\documentclass[cameraready]{Interspeech}
\usepackage{booktabs}
\usepackage{multirow}
\usepackage{shuffle}
\usepackage{amssymb}
\usepackage{xcolor}
\usepackage{cite}
\usepackage{tikz}
\usetikzlibrary{arrows.meta,automata,positioning}
\usepackage{makecell}

\usetikzlibrary{quotes}
\usepackage{ifthen}
\usepackage[linesnumbered,ruled,vlined]{algorithm2e}
\usepackage{stmaryrd}
\usetikzlibrary{automata, positioning}
\usepackage{subcaption}
\usepackage{comment}

\newcommand{\bigshuffle}{\mathop{\scalebox{1.8}{$\shuffle$}}\displaylimits}

\DeclareMathOperator*{\argmax}{arg\,max}

\title{Modeling Overlapped Speech with Shuffles}
\author[affiliation={1,2}, orcid=0000-0002-5423-7754, ]{Matthew}{Wiesner}
\author[affiliation={3}, orcid=0000-0002-5358-1844, ]{Samuele}{Cornell}
\author[affiliation={4}, orcid=0009-0000-4958-202X, ]{Alexander}{Polok}
\author[affiliation={2}, orcid=0000-0003-4512-0471, ]{Lucas}{Ondel Yang}
\author[affiliation={4}, orcid=0000-0002-4951-5908, ]{Lukáš~Burget}{}
\author[affiliation={1}, orcid=0000-0001-5976-0897, ]{Sanjeev}{Khudanpur}

\address{
    $^1$ Johns Hopkins University, USA,
    $^2$ LISN, CNRS, France \\
    $^3$ Carnegie Mellon University, USA,   
    $^4$ Brno University of Technology, Czechia
}

\email{\{wiesner,khudanpur\}@jhu.edu, lucas.ondel@cnrs.fr, samuele.cornell@ieee.org, \{ipoloka,burget\}@fit.vutbr.cz}


\begin{document}

\maketitle

\begin{abstract}
We propose to model parallel streams of data, such as overlapped speech, using shuffles. Specifically, this paper shows how the shuffle product and partial order finite-state automata (FSAs) can be used for alignment and speaker-attributed transcription of overlapped speech. We train using the total score on these FSAs as a loss function, marginalizing over all possible serializations of overlapping sequences at subword, word, and phrase levels. To reduce graph size, we impose temporal constraints by constructing partial order FSAs. We address speaker attribution by modeling (token, speaker) tuples directly. Viterbi alignment through the shuffle product FSA directly enables one-pass alignment. We evaluate performance on synthetic LibriSpeech overlaps. To our knowledge, this is the first algorithm that enables single-pass alignment of multi-talker recordings. All algorithms are implemented using k2 / Icefall. 

\end{abstract}

\section{Introduction}
\label{sec:intro}
This work demonstrates a way to extend models of concurrency to the problem of end-to-end (e2e) speaker-attributed alignment and transcription of overlapped speech. Two commonly used models for representing concurrent processes are partial orders~\cite{pratt1986modeling, godefroid1996partial}, and shuffle products~\cite{gischer1981shuffle, restivo2015shuffle}. The shuffle product, which generates all possible interleavings of sequences that preserve the internal order of each sequence, has been used to represent the space of possible serializations of concurrent events\cite{restivo2015shuffle}. Similarly, one can represent a partial order on a subset of sequences as the set of all admissible serializations.

Overlapped speech presents a natural instance of this problem. We model multi-talker transcription as the recovery of interleaved sequences of discrete linguistic units: subwords, words, or phrases, observed through a shared acoustic channel. Since the exact temporal ordering of tokens across speakers is unknown, we show that the Connectionist Temporal Classification (CTC)~\cite{graves2006connectionist} objective can be extended to this setting by replacing the standard supervision sequence with a finite-state automaton (FSA) representing the shuffle product of utterance transcripts. The forward algorithm on this FSA marginalizes over all valid interleavings, in the same way that standard CTC marginalizes over all valid frame-to-token alignments. To keep the approach tractable, we introduce partial order constraints that prune the shuffle FSA using approximate token timing, yielding a principled trade-off between flexibility and computational cost. We show that other approaches for multi-talker ASR such as utterance-level Serialized Output Training (SOT)~\cite{kanda2020serialized}, token-level SOT~\cite{kanda2022streaming}, and Speaker-Distinguishable SD-CTC~\cite{sdctc} can be understood as special cases or relaxations of this framework.

A benefit of this formulation is that it produces aligned, speaker-attributed transcripts even in the presence of overlapped speech. 
The Viterbi path through the shuffle product FSA directly provides a single-pass alignment of overlapped speech, which, to our-knowledge, has not previously been done. We extend this framework to also model speaker attribution by augmenting the CTC output to (token, speaker) tuples, and propose both factored and joint models of this distribution. We describe methods for 1-pass speaker-attributed decoding for multi-talker speech using CTC argmax decoding and a target-speaker decoding method that enables standard WFST language model integration.

These contributions open up new research possibilities particularly relevant for processing in-the-wild multi-talker data e.g. from webvideos~\cite{li2023yodas}, where  per-speaker audio streams are unavailable and existing annotations---such as subtitles or approximate transcripts with speakers turn---are often temporally misaligned especially when speakers overlap.

More broadly, although we demonstrate our approach using CTC for simplicity, it should be readily extensible to other objective functions that can be represented by weighted finite-state transducers (WFSTs)~\cite{mohri2002weighted}, including the transducer~\cite{graves2012sequence} and hidden Markov models (HMMs)~\cite{rabiner2002tutorial}. The formulation itself is also not specific to speech: it applies whenever multiple independent processes emit sequences that are observed interleaved through a single channel without a known ordering, as in polyphonic music transcription, multi-agent activity recognition, or multi-event detection.

This paper first introduces the shuffle product and shows how partial order constraints can be included to make the graph size manageable, even for long sequences of overlapped speech. Then we show how these core techniques can be applied to the problem of training, decoding, and aligning overlapped speech. In the process we show how prior work including tSOT, SOT, and SD-CTC relate to the shuffle product. Finally, we explore our proposed training, decoding, and alignment methods on synthetic LibriSpeech overlaps.

\section{Related Work}
\label{sec:related}
Several approaches have been proposed for handling overlapped speech in multi-talker ASR. Early work used Permutation Invariant Training (PIT)~\cite{yu2017permutation, kolbaek2017multitalker}, where multiple output heads share an encoder and each branch is trained with a CTC loss~\cite{yu2017recognizing, seki2018purely, chang2020end}. However, PIT requires fixing the maximum number of speakers a priori and scales poorly, as each additional speaker introduces a new branch and increases the number of loss evaluations combinatorially.

Serialized Output Training (SOT)~\cite{kanda2020serialized} sidesteps this by concatenating multi-speaker transcripts into a single output sequence, enabling standard sequence-to-sequence training. The utterance-level FIFO ordering it imposes, however, may not reflect the temporal dynamics of the signal. In token-level SOT (tSOT)~\cite{kanda2022streaming}, tokens are ordered according to their start times, so they are temporally aligned, but this requires accurate word-level alignments which do not generally exist for in-the-wild data where overlapped speech must be handled directly.

More recently, Speaker-Distinguishable CTC (SD-CTC)~\cite{sdctc} extends CTC to multiple speakers by introducing speaker-specific blank symbols and an auxiliary speaker output. We show in Section~\ref{sec:sdctc_relationship} that SD-CTC can be interpreted as a relaxation of our framework and that token-level SOT, and SD-CTC can all be viewed as special cases within our formulation (Sections~\ref{sec:partial_orders}), providing a unifying perspective on these seemingly disparate approaches. Recent models such as Serialized Output Prompting \cite{shi2025serialized} and GEncSep \cite{GEncSep} also use CTC to produce speaker-ordered representations of overlapped speech, which are then passed to a downstream attention decoder or LLM. Our method could similarly replace the speaker-serialized CTC component used in those approaches. 

Our work builds on a line of research extending CTC-like marginalization to weakly supervised settings. Semi-supervised MMI training~\cite{manohar2015semi, manohar2018semi} uses ASR lattices as supervision, while Bypass Temporal Classification~\cite{gao2023bypass}, OTC~\cite{gao2023learning}, and Star-CTC~\cite{pratap2022star} extend CTC to work directly on errorful transcripts without prior lattice construction from existing ASR systems. \cite{klejch2021deciphering} even use a full language model for phoneme decipherment. These ideas have also been extended to transducer models~\cite{bataev2025rnn, laptev2023powerful}.
These methods show that composing the CTC topology with flexible supervision graphs that accept all admissible sequences subject to annotation constraints enables training in weakly supervised settings. We extend this concept to multi-talker transcripts: since the exact interleaving of tokens across speakers is unknown, we compose the CTC topology with shuffle product FSAs that encode all valid interleavings.

\section{Method}
\subsection{Shuffles for Multi-talker Speech}
\label{sec:shuffles}

The shuffle product, which generates all possible interleavings of sequences while preserving the internal order of each, is a well-studied construct in formal language theory and concurrency modeling~\cite{gischer1981shuffle, restivo2015shuffle, pratt1986modeling}. In the \emph{poset} (partially ordered sets) literature~\cite{davey2002introduction}, these interleavings are called linear extensions, or linearizations. We use the term serialization here to emphasize the connection with serialized output training commonly used in multi-talker ASR.

Regular languages are closed under the shuffle operator, and the resulting language can be compactly represented as a finite-state automaton (FSA). To our knowledge, the shuffle product has not previously been used in multi-talker speech recognition. We show that it provides an elegant framework for modeling overlapped speech, where the interleaving of tokens from concurrent speakers is unknown.

Let $y, z$ be two sequences and let $y \shuffle z$ denote the shuffle between them. It is the set of all possible riffle shuffles of the two sequences. For example, if $y = \mbox{abc}, z = \color{red}\mbox{xy}$, then
\begin{align*}
    y\,\shuffle\,z \color{black} = \{&\color{black}\mbox{abc}\color{red}\mbox{xy}\color{black},
    \color{black}\mbox{ab}\color{red}\mbox{xy}\color{black}\mbox{c},
    \mbox{a}\color{red}\mbox{xy}\color{black}\mbox{bc},
    \color{red}\mbox{xy}\color{black}\mbox{abc},
    \color{black}\mbox{ab}\color{red}\mbox{x}\color{black}\mbox{c}\color{red}\mbox{y}\color{black}, \color{black}\mbox{a}\color{red}\mbox{x}\color{black}\mbox{bc}\color{red}\mbox{y}\color{black},\\
    &\color{red}\mbox{x}\color{black}\mbox{abc}\color{red}\mbox{y}\color{black}, 
    \mbox{a}\color{red}\mbox{x}\color{black}\mbox{b}\color{red}\mbox{y}\color{black}\mbox{c}, 
    \color{red}\mbox{x}\color{black}\mbox{ab}\color{red}\mbox{y}\color{black}\mbox{c}, \color{red}\mbox{x}\color{black}\mbox{a}\color{red}\mbox{y}\color{black}\mbox{bc}\}.
\end{align*}
The notion of shuffle can be extended to sets of sequences. Let $L_1, L_2$ be two such sets. Then $L_1 \shuffle L_2 = \bigcup_{y \in L_1} \bigcup_{z \in L_2} y \shuffle z$ with $\emptyset \shuffle \{y\} = \{y\}$. Defining the shuffle of a sequence with a set enables shuffling more than two sequences together via iterative pairwise shuffling, which returns the same set regardless of the shuffling order because the shuffle is both associative and commutative~\cite{restivo2015shuffle}. We use $\bigshuffle_{y \in L} y$ to represent the iterated shuffle product of sequences in $L$.

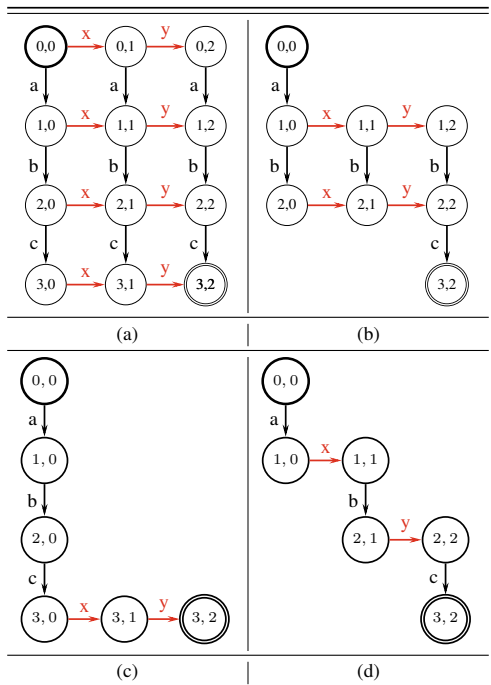
\begin{figure}[t]
  \centering
  \resizebox{0.8\columnwidth}{!}{%
  \begin{tabular}{c|c}
    \toprule
        \midrule
        \begin{tikzpicture}[>=Stealth, >={Stealth[length=2mm, width=1mm]}]

  \definecolor{myred}{RGB}{220,50,32}
  \def\sp{1.3} 
  
  \foreach \i in {0,1,2,3}{
    \foreach \j in {0,1,2}{
      \ifnum\i=0
        \ifnum\j=0
          \node[state, thick, line width=1.2pt, minimum size=12pt] (s\i\j) at (\j*\sp,-\i*\sp) {\scriptsize 0,0};
        \else
          \node[state, minimum size=12pt] (s\i\j) at (\j*\sp,-\i*\sp) {\scriptsize \i,\j};
        \fi
      \else
        \node[state, minimum size=12pt] (s\i\j) at (\j*\sp,-\i*\sp) {\scriptsize \i,\j};
      \fi
    }
  }

  \node[accepting,state,minimum size=12pt] at (2*\sp,-3*\sp) (s32) {\scriptsize 3,2};

  \draw[->, thick] (s00) -- node[left]  {a} (s10);
  \draw[->, thick] (s10) -- node[left]  {b} (s20);
  \draw[->, thick] (s20) -- node[left]  {c} (s30);

  \draw[->, thick] (s01) -- node[left]  {a} (s11);
  \draw[->, thick] (s11) -- node[left]  {b} (s21);
  \draw[->, thick] (s21) -- node[left]  {c} (s31);

  \draw[->, thick] (s02) -- node[left]  {a} (s12);
  \draw[->, thick] (s12) -- node[left]  {b} (s22);
  \draw[->, thick] (s22) -- node[left]  {c} (s32);

  \draw[->, thick, myred] (s00) -- node[above] {x} (s01);
  \draw[->, thick, myred] (s01) -- node[above] {y} (s02);

  \draw[->, thick, myred] (s10) -- node[above] {x} (s11);
  \draw[->, thick, myred] (s11) -- node[above] {y} (s12);

  \draw[->, thick, myred] (s20) -- node[above] {x} (s21);
  \draw[->, thick, myred] (s21) -- node[above] {y} (s22);

  \draw[->, thick, myred] (s30) -- node[above] {x} (s31);
  \draw[->, thick, myred] (s31) -- node[above] {y} (s32);

\end{tikzpicture} &
        \begin{tikzpicture}[>=Stealth, >={Stealth[length=2mm, width=1mm]}]

\definecolor{myred}{RGB}{220,50,32}
\def\sp{1.3} 

\node[state, thick, line width=1.2pt, minimum size=12pt] (s00) at (0*\sp, -0*\sp) {\scriptsize 0,0};
\node[state, minimum size=12pt] (s10) at (0*\sp, -1*\sp) {\scriptsize 1,0};
\node[state, minimum size=12pt] (s20) at (0*\sp, -2*\sp) {\scriptsize 2,0};

\node[state, minimum size=12pt] (s11) at (1*\sp, -1*\sp) {\scriptsize 1,1};
\node[state, minimum size=12pt] (s21) at (1*\sp, -2*\sp) {\scriptsize 2,1};

\node[state, minimum size=12pt] (s12) at (2*\sp, -1*\sp) {\scriptsize 1,2};
\node[state, minimum size=12pt] (s22) at (2*\sp, -2*\sp) {\scriptsize 2,2};
\node[accepting,state,minimum size=12pt] (s32) at (2*\sp, -3*\sp) {\scriptsize 3,2};


\draw[->, thick] (s00) -- node[left]  {\normalsize a} (s10);
\draw[->, thick] (s10) -- node[left]  {\normalsize b} (s20);

\draw[->, thick] (s11) -- node[left]  {\normalsize b} (s21);
\draw[->, thick] (s12) -- node[left]  {\normalsize b} (s22);
\draw[->, thick] (s22) -- node[left]  {\normalsize c} (s32);


\draw[->, thick, myred] (s10) -- node[above] {\normalsize x} (s11);
\draw[->, thick, myred] (s11) -- node[above] {\normalsize y} (s12);

\draw[->, thick, myred] (s20) -- node[above] {\normalsize x} (s21);
\draw[->, thick, myred] (s21) -- node[above] {\normalsize y} (s22);

\end{tikzpicture} \\
        \midrule
        (a) & (b) \\
        \midrule
        \begin{tikzpicture}[>=Stealth, >={Stealth[length=2mm, width=1mm]}]
  \definecolor{myred}{RGB}{220,50,32}
  \def\sp{1.3} 
  
  \node[state, thick, minimum size=12pt, line width= 1.2pt] (s00) at (0*\sp,0)        {\scriptsize $0,0$};
  \node[state, thick, minimum size=12pt] (s10) at (0*\sp, -1*\sp)  {\scriptsize $1,0$};
  \node[state, thick, minimum size=12pt] (s20) at (0*\sp, -2*\sp)  {\scriptsize $2,0$};
  \node[state, thick, minimum size=12pt] (s30) at (0*\sp, -3*\sp)  {\scriptsize $3,0$};
  \node[state, thick, minimum size=12pt] (s31) at (1*\sp, -3*\sp)  {\scriptsize $3,1$};
  \node[state, accepting,thick, minimum size=12pt] (s32) at (2*\sp,-3*\sp) {\scriptsize $3,2$};

  \draw[->, thick]        (s00) -- node[left]  {a} (s10);
  \draw[->, thick]        (s10) -- node[left]  {b} (s20);
  \draw[->, thick]        (s20) -- node[left]  {c} (s30);

  \draw[->, thick, myred] (s30) -- node[above] {x} (s31);
  \draw[->, thick, myred] (s31) -- node[above] {y} (s32);

\end{tikzpicture} & \begin{tikzpicture}[>=Stealth, >={Stealth[length=2mm, width=1mm]}]
  \definecolor{myred}{RGB}{220,50,32}
  \def\sp{1.3}

  \node[state, thick, minimum size=12pt, line width=1.2pt] (s00) at (0*\sp,0)        {\scriptsize $0,0$};
  \node[state, thick, minimum size=12pt] (s10) at (0*\sp, -1*\sp)  {\scriptsize $1,0$};
  \node[state, thick, minimum size=12pt] (s11) at (1*\sp, -1*\sp)  {\scriptsize $1,1$};
  \node[state, thick, minimum size=12pt] (s21) at (1*\sp, -2*\sp)  {\scriptsize $2,1$};
  \node[state, thick, minimum size=12pt] (s22) at (2*\sp, -2*\sp)  {\scriptsize $2,2$};
  \node[state, accepting, thick, minimum size=12pt] (s32) at (2*\sp, -3*\sp) {\scriptsize $3,2$};

  \draw[->, thick]         (s00) -- node[left]  {a} (s10);
  \draw[->, thick, myred]  (s10) -- node[above] {x} (s11);
  \draw[->, thick]         (s11) -- node[left]  {b} (s21);
  \draw[->, thick, myred]  (s21) -- node[above] {y} (s22);
  \draw[->, thick]         (s22) -- node[left]  {c} (s32);

\end{tikzpicture} \\
        \midrule
        (c) & (d) \\
        \bottomrule
    \end{tabular}
}
  \caption{FSAs representing different utterance group serializations. Tuples on states represent indices into the two serialized sequences. Colors represent unique sources. (a) The shuffle product FSA between two sequences $y_1 = abc, y_2 = \color{red}xy$. (b) The pruned shuffle FSA obtainable from application of partial order constraints (see Section \ref{sec:partial_orders}). (c) The subgraph corresponding to utterance-level serialized output training (SOT). (d) The subgraph corresponding to token-level SOT, assuming token order $a \leq \color{red}x\color{black} \leq b \leq \color{red}y\color{black} \leq c$.}
  \label{fig:shuffle_product_fsa}
\end{figure}

For two sequences of length $M, N$, each element in the shuffle has length $M + N$. The number of elements in the shuffle is $\binom{M+N}{N}$. An FSA, accepting all elements of the shuffle product can be constructed from a set of sequences.  Figure~\ref{fig:shuffle_product_fsa}(a) shows the FSA shuffle product for two example sequences. Consider the shuffle of $k$, length-$N$ sequences. The space complexity of the graph is $O\left(kN^k\right)$ ($N^k$ states for the hypercube, each with $k$ outgoing arcs). Scaling to more than 2 overlapping sequences, especially longer ones, therefore requires strategies to limit graph size.

\subsection{Partial Orders and Serialization}
\label{sec:partial_orders}

Partial order constraints are one way of limiting the size of the shuffle product FSA. A partial order is a precedence relationship on a subset of elements. Elements in this subset are called comparable elements~\cite{davey2002introduction}, while those not in the subset cannot be ordered. The shuffle product itself encodes a partial order since it requires that the internal order of each shuffled sequence be preserved without imposing any ordering between elements of different sequences. Intra-sequence elements are therefore comparable while inter-sequence elements are not.

The precedence relationship of the shuffle product can be extended to inter-sequence tokens by using approximate token start times to discard unlikely interleavings, which enables further pruning of states in the shuffle graph. Token start times can be approximated by linearly interpolating between utterance start and end times (see Section~\ref{sec:serialization}).

Let $t_i, t_j \in \mathbb{R}$ be the start times of the $i^{th}$ and $j^{th}$ tokens in sequences $y$ and $z$ respectively, and let $\kappa$ be a threshold controlling the amount of uncertainty allowed  in the temporal ordering, i.e., a ``collar.'' We can then define a partial order over tokens, $y_i \preceq z_j$ if $t_i < t_j - \kappa$, meaning $y_i$ must appear before $z_j$ in any valid serialization. Tokens whose start times fall within $\kappa$ of each other are left unordered, allowing for all possible interleavings between them. The resulting FSA is obtained by pruning the full shuffle product FSA to retain only paths consistent with these temporal constraints via a breadth-first search. \footnote{We defined the partial order on tokens for ease of exposition, however, defining the partial order over the states of the shuffle graph is more appropriate and resolves some ambiguity regarding tokens with the same label in different sequences. All states in the shuffle graph that do not obey the imposed partial order are pruned.}

Figure~\ref{fig:shuffle_product_fsa}(b) depicts the pruned shuffled graph from (a) after imposing additional inter-sequence partial order constraints. Given the ordering constraints $\mbox{a} \preceq \mbox{b} \preceq \mbox{c}$, $\color{red}\mbox{x}\color{black} \preceq \color{red}\mbox{y}$, $\color{black}\mbox{a} \preceq \color{red}\mbox{x}$, and $\color{red}\mbox{y}\color{black} \preceq \mbox{c}$, only tokens $\mbox{b}$, $\color{red}\mbox{x}$, and $\color{red}\mbox{y}$ remain mutually unordered, significantly reducing the number of valid paths. The forward algorithm on this pruned FSA marginalizes over the remaining valid interleavings.

\subsubsection{Relationship to existing serialization schemes}
\label{sec:serialization}

Let $y = \{y_1, \ldots, y_M\}$, $z = \{z_1, \ldots, y_N\}$ represent utterance transcripts of $M$ and $N$ tokens respectively, with known start times, $b_y \leq b_z$, and end times $e_y, e_z$. Commonly used serializations can be viewed as distinct partial order constraints.
\begin{enumerate}
    \item \textbf{Utterance-level SOT} imposes a total order at the speaker level, concatenating all tokens of the first utterance before the second: $\{y_1, \ldots, y_M, z_1, \ldots, z_N\}$ (Figure~\ref{fig:shuffle_product_fsa}(c)). This is a serialization strategy that constrains the order of speakers.

    \item \textbf{Token-level SOT}, such as in~\cite{kanda2022streaming}, orders tokens by start time, which requires token alignments. Let $t_{y_i}$, be the start time of the $i^{th}$ token in the sequence $y$. We approximately infer these start times using the heuristic
    \begin{equation}
        t_{y_i} = b_y + i \cdot \frac{e_y - b_y}{M}.
        \label{eq:token_time_heuristic}
    \end{equation}
    This enforces a single deterministic interleaving (Figure~\ref{fig:shuffle_product_fsa}(d)) and corresponds to $\kappa = 0$ in our framework.

    \item \textbf{Partial ordering with a collar} allows any interleaving for which token start times are within $\kappa$ of each other. Intuitively, this is analogous to the time-constrained permutation used in tcpWER~\cite{mimower}: just as tcpWER allows flexible speaker-to-reference matching within a time collar during evaluation, our partial order allows flexible token interleaving within a collar during training.

    \item \textbf{Full shuffle} is recovered when $\kappa \to \infty$, encoding all possible interleavings with no temporal constraints.
\end{enumerate}
Sweeping $\kappa$ from $0$ to $\infty$ provides a continuum from token-level SOT to the full shuffle, where the appropriate collar depends on the accuracy of the available token-level timing supervision.

\subsection{CTC with Shuffles}
\label{sec:ctc_shuffles}
%
We now show how to train acoustic models using these shuffle-graph supervisions in conjunction with CTC.
The loss used in CTC is a particular instance of the forward algorithm computed on an FSA representing all possible alignments of a ground-truth output sequence to the input acoustic sequence. Because the exact alignment is unknown, the forward algorithm marginalizes over all admissible alignments.
We extend this idea to multi-talker transcripts without known token-level time information. Since we do not know the exact order of token sequences among the utterances, we use shuffles and partial orders to construct the set of all possible interleavings. In the same way that CTC marginalizes over all valid alignments of input frames to output tokens, the forward algorithm on the shuffle product FSA marginalizes over all possible ways of interleaving training utterances.

\subsubsection{Extending CTC to Utterance Groups with Shuffles}
Let $x_t$ be a sample of speech at time $t$ and $\mathbf{x} = \left(x_1, \ldots, x_T\right)$ be a sequence of such samples, i.e., an audio input. Let $a = a_1 \ldots a_T$, be an alignment of the speech frames to the transcript, where $a_t$ is either an element from a vocabulary, $\mathcal{V}$ or no-speech, also commonly called blank and denoted by $\oslash$. The transcript can be recovered from an alignment by collapsing repeated symbols and removing any $\oslash$ symbols. Let $\beta\left(\right)$ represent this token collapsing function, and let $\beta^{-1}\left(y\right) = \{a |\, \beta\left(a\right) = y\}$ be the set of all admissible alignments for transcript $y$. CTC models the probability of a transcript, $y$, by marginalizing over all alignments, $a$, and assuming output symbol probabilities are independent conditioned on the input, i.e.,
\begin{equation}
    p\left(y | \mathbf{x}\right) = \sum_{a \in \beta^{-1}\left(y\right)} p\left(a | \mathbf{x}\right) = \sum_{a \in \beta^{-1}\left(y\right)} \prod_{t=1}^T p\left(a_t | \mathbf{x}\right).
\end{equation}

Consider a set of $k$ utterances, $\mathcal{G} = \{u^1, \ldots, u^k\}$, sometimes called an utterance group in the literature~\cite{kanda21_interspeech} associated with the audio input, $\mathbf{x}$.  Let $\mathcal{S} = \{1, \ldots, s_{max}\}$ be an ordering of up to $s_{max}$ speakers in $\mathbf{x}$, e.g., the order in which speakers first appear, and let $y$ be a token sequence. Finally, let $t_b, t_e$ be utterance beginning and end times. Then $u^i = \left(t_b, t_e, s, y\right)$. To denote the start, $t_b$, or speaker, $s$, of the utterance we use $u^i\left[t_b\right]$, $u^i\left[s\right]$ etc.. Let $\mathcal{Y} = \bigshuffle\limits_{u \in \mathcal{G}} u\left[y\right]$ be the shuffle product of the token sequences in $\mathcal{G}$ and let $\beta^{-1}\left(\mathcal{Y}\right)$ be the set of alignments, to frames in $\mathbf{x}$, of elements of the shuffle product corresponding to the utterance group, $\mathcal{G}$. We want to maximize the probability of the utterance group transcripts.
\begin{align}
    p\left(\mathcal{G} | \mathbf{x}\right) &= \sum_{y \in \mathcal{Y}} p\left(y | \mathbf{x}\right) = \sum_{y \in \mathcal{Y}} \sum_{a \in \beta^{-1}\left(y\right)} p\left(a | \mathbf{x}\right) \\
    &= \sum_{a \in \beta^{-1}\left(\mathcal{Y}\right)} \prod_{t=1}^T p\left(a_t | \mathbf{x}\right). \label{eq:ctc_loss}
\end{align}

This looks exactly like the probability of a transcript as modeled by CTC, but we have expanded the set of possible alignments to include multi-talker FSAs using shuffles. We refer to the negative log of this quantity as the shuffle loss. It can be implemented using the FSAs shown in Figure \ref{fig:shuffle_product_fsa}. Let $U\left(\mathbf{x}\right)$ be the dense FSA that encodes the probabilities $p\left(a_t | \mathbf{x}\right) \, \forall\, t$, of each possible output token, $a_t$, at each moment in time, $t$. Let $G\left(\mathcal{Y}\right)$ represent the supervision graph, i.e., the shuffle FSA. Then in the log-semiring

\begin{equation}
\log{p\left(\mathcal{G} | \mathbf{x}\right)} = \llbracket U\left(\mathbf{x}\right) \circ \left(T \circ G\left(\mathcal{Y}\right)\right)\rrbracket. \label{eq:forward_score}
\end{equation}

$\circ$ is the FST composition, and $\llbracket A \rrbracket$ is the forward score (sum of products) over all paths in the graph, $A$. See \cite{mohri2009weighted} for more background on FSA algorithms.

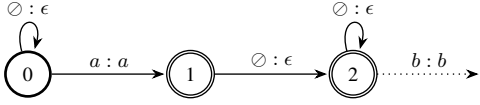
\begin{figure}
    \centering
    \resizebox{0.8\columnwidth}{!}{%
    \begin{tikzpicture}[->,>=Stealth,shorten >=1pt,auto,node distance=1.8cm,semithick]
      \tikzset{
        state/.style={circle,draw,minimum size=20pt,inner sep=1pt},
        start/.style={state, very thick}
      }
    
      \node[start] (s0) {0};
      \node[state, accepting] (s1) [right=of s0] {1};
      \node[state, accepting] (s2) [right=of s1] {2};
      \path 
            (s0) edge[loop above] node {$\oslash:\epsilon$} (s0)
            (s0) edge node {$a:a$} (s1)
            (s1) edge node {$\oslash:\epsilon$} (s2)
            (s2) edge[loop above] node {$\oslash:\epsilon$} (s2)
            (s2) edge[dotted] node {$b:b$} ++(2cm,0);
    \end{tikzpicture}
    }
    \caption{The compact selfless label topology from \cite{laptev2022ctc}. Note that, different from the traditional CTC-topology, it forces tokens to appear on exactly one frame, and forces $\oslash$ between every token in the transcript, i.e., between ``a'' and a subsequent ``b''.}
    \label{fig:ctc_compact}
\end{figure}

Prior work \cite{zeyer2021does} has shown that allowing the $\oslash$ symbol between any tokens results in peaky label predictions. \cite{laptev2022ctc} leverage this behavior to reduce the size of decoding graphs by exploring alternative CTC topologies, $T$. Since we are modeling overlapped speech as a sequence of discrete, interleaved events (word / phoneme start-times), and we are not aiming for frame-level classification, the peaky behavior of CTC is actually desirable. Using a reduced topology that exaggerates the peaky behavior of CTC also helps to reduce the size of $T \circ G\left(\mathcal{Y}\right)$, which is critical since $\mathcal{Y}$ can be large. We therefore use the compact selfless topology of \cite{laptev2022ctc} (see Figure \ref{fig:ctc_compact}).

\subsection{Speaker Attribution with Shuffles}
\label{sec:speaker_attribution}
We now discuss how to attribute tokens to speakers. To each token, $y_t$, in each element in the shuffle product, $y \in \mathcal{Y}$, we will associate a speaker tag, $s_t \in \mathcal{S}$, from some predefined set of speaker labels. In other words, the shuffle elements, $y = \left( \left(y_1, s_1\right), \left(y_2, s_2\right), \ldots, \left(y_T, s_T\right) \right) \in \mathcal{Y} \times \mathcal{S}$, are now interleaved tuples. Let $\mathcal{Y}^{\prime}$ denote this speaker augmented shuffle product. In CTC-like objective functions, e.g., Equation \ref{eq:ctc_loss}, we can replace $a_t$ so that instead of tokens, $a_t \in \mathcal{V} \bigcup \{\oslash\}$, they are now tuples or blank symbols, $a_t^\prime \in \left(\mathcal{V} \times \mathcal{S}\right) \bigcup \{\oslash\}$, e.g.,
\begin{equation*}
    a^\prime = \left(\oslash, \oslash, \left(\upsilon_3, \sigma_3\right), \oslash, \ldots, \oslash, \left(\upsilon_{T-1}, \sigma_{T-1}\right), \oslash\right).
\end{equation*}
The supervision graph, $G\left(\mathcal{Y}\right)$, in Equation \ref{eq:forward_score}, changes correspondingly to $G\left(\mathcal{Y}^\prime\right)$. Let $\phi\left(\mathbf{x}\right)$ represent the outputs from a neural network, $\phi\left(\cdot\right)$, such as WavLM \cite{chen2022wavlm}. Let $W_y, W_s$ be linear projections from the neural network output embedding dimension to the output vocabulary and number of speakers respectively. Let $\bar{\oslash}$ be the shorthand for $a_t^\prime \neq \oslash$. We can model the joint, frame-level probability by computing joint scores for speakers and tokens using a factored joint model as in speaker distinguishable CTC \cite{sdctc}:
\begin{align*}
    &p_{a_t^\prime}\left(a_t^\prime | \mathbf{x}\right)
    = \begin{cases} p_{\upsilon_t}\left(\oslash | \mathbf{x}\right); & a_t^\prime = \oslash \\
    p_{\upsilon_t}\left(\upsilon_t | \mathbf{x} \right) p_{\sigma_t}\left(\sigma_t | \mathbf{x}, \bar{\oslash}\right); & a_t^\prime \neq \oslash
    \end{cases}
    \label{eq:factored_attribution}
    \\
    &p_{\upsilon_t}\left(\upsilon_t | \mathbf{x}\right) = \mbox{Softmax} \left(W_y \phi\left(\mathbf{x}\right)\right)_{\upsilon_t}^t \\
    &p_{\sigma_t}\left(\sigma_t | \mathbf{x}, \bar{\oslash}\right) = \mbox{Softmax} \left(W_s \phi\left(\mathbf{x}\right)\right)_{\sigma_t}^t.
\end{align*}

The way to interpret $p_{\sigma_t}\left(\sigma_t | \mathbf{x}, \bar{\oslash}\right)$, especially in the context of overlapped speech, is the probability of speaker $\sigma_t$ given that time, $t$, is the start of a non-$\oslash$ token. Note that we have overloaded $\upsilon_t$ to be simultaneously the index of a symbol and its representation.

Alternatively, we can directly normalize over scores for all speaker-token pairs,
\begin{align*}
    f\left(\oslash | \mathbf{x}\right) &= \left(W_y \phi\left(\mathbf{x}\right) \right)_\oslash^t, \\
    f\left(\upsilon_t, \sigma_t | \mathbf{x}, \bar{\oslash}\right) &= \left(W_y \phi\left(\mathbf{x}\right)\right)_{\upsilon_t}^t + \left(W_s \phi\left(\mathbf{x}\right)\right)_{\sigma_t}^t,
\end{align*}
and unrolling into a vector the scores $f\left(\upsilon_t, \sigma_t | \mathbf{x}, \bar{\oslash}\right)$ of all speaker, token pairs lets us define the joint probability, 
\begin{equation*}
p\left(a_t^\prime | \mathbf{x}\right) = \mbox{Softmax}\left(\left[f\left(\oslash | \mathbf{x}\right),  \{f\left(\upsilon_t, \sigma_t | \mathbf{x}\right)\}_{\upsilon, \sigma \in \mathcal{V} \times \mathcal{S}}\right]\right).
\end{equation*}
We refer to this as the direct joint model.

Speaker labels can be produced as in \cite{raj2024speaker, parksortformer}, where the speaker label is assigned by the order in which speakers appear in the conversation, but other heuristics such as total amount of spoken time could be used.

\subsection{Multi-talker Decoding}
CTC decoding is commonly improved by fusion with a language model. In the WFST framework this is done by composing a graph representation of the neural network output probabilities with FSTs representing n-gram language models and perhaps a pronunciation, or byte-pair encoding (BPE)-unit lexicon. Unfortunately, we do not have shuffle n-gram language models. The space of all possible shuffles is also too large to represent as a graph. Instead, we propose two methods to perform multi-talker decoding with shuffles.

\subsubsection{1-pass Decoding}
\label{sec:1-pass_decoding}
The simplest approach is 1-pass greedy CTC decoding, which simultaneously transcribes all speakers. Since the model predicts a (BPE unit, speaker label) tuple at each frame, the optimal token sequence and speaker assignment can be recovered directly:
\begin{align*}
    a_t^* &= \operatorname*{argmax}\limits_{a_t \in (\mathcal{S} \times \mathcal{V}) \bigcup \oslash} p\left(a_t | \mathbf{x}\right) \\
    y^* &= \beta\left(a_1^*\, \vert\vert\, \cdots \vert\vert\, a_N^*\right),
\end{align*}
where $\vert\vert$ represents concatenation and $\beta\left(\cdot\right)$ is the CTC-collapse function that removes repeated and then blank symbols. The only remaining issue is how to recover word-level transcripts from this token sequence. Decoded tokens, $(\upsilon, \sigma)$, with the same speaker label, $\sigma$, are concatenated and detokenized according to the BPE model. 

Some evaluation metrics (see Section \ref{sec:metrics}) additionally require timing information. We take the start time of each token to be the frame at which it was decoded. Its duration is the number of frames until the same speaker's next token. If the gap between consecutive tokens from the same speaker exceeds a threshold (we use 0.5\,s), we treat it as an utterance boundary and set the final token's duration to the average duration of the other tokens in that utterance.

\subsubsection{N-pass Decoding}
\label{sec:tgt_speaker_decoding}
Language model integration requires disentangling the output sequences so that each is a single speaker utterance. A multi-pass decoding method can be applied whereby the neural network outputs are repeatedly masked and passed through a CTC decoding graph, $TLG$. Let $\oslash_\sigma$, be a new speaker specific blank probability, as used in SD-CTC \cite{sdctc}. The following target speaker network outputs are used for decoding target speaker $\sigma$:
\begin{align}
    p\left(\oslash_\sigma | \mathbf{x}\right) &= p_{\upsilon_t}\left(\oslash | \mathbf{x} \right) +  \sum_{\rho \neq \sigma}p_{\upsilon_t}\left(\upsilon_t | \mathbf{x} \right) p_{\sigma_t}\left(\rho | \mathbf{x}, \bar{\oslash}\right)\\ 
    p_t\left(\upsilon, \sigma | \mathbf{x}, \bar{\oslash}\right) &= p_{\upsilon_t}\left(\upsilon_t | \mathbf{x} \right) p_{\sigma_t}\left(\sigma | \mathbf{x}, \bar{\oslash}\right)
\end{align}

In other words, the probability mass corresponding to other speakers is transferred to $\oslash$. Let the WFSA, $U_\sigma\left(\mathbf{x}\right)$, be the dense FSA representing these target-speaker frame-level probabilities. Then decoding can be performed as 
\begin{align*}
    \hat{w}_\sigma &= \mbox{Proj}_w\left(\mbox{ShortestPath}\left(U_\sigma\left(\mathbf{x}\right) \circ \left(T \circ L \circ G \right)\right)\right) \\
    \mathcal{G} &= \{ \hat{w}_\sigma \,\vert\, \sigma \in 1, \ldots, |\mathcal{S}|\},
\end{align*}
where $\mbox{Proj}_w$ is the FST projection onto the output labels. Timing information is recovered in the same way as in the 1-pass decoding algorithm.
\subsection{Relationship to Speaker Distinguishable CTC}
\label{sec:sdctc_relationship}
The $N$-pass decoding method described in the previous section is closely related to the \emph{training} procedure used in Speaker-Distinguishable CTC (SD-CTC) \cite{sdctc}. We make this relationship explicit here. Let \begin{align}
y_{\sigma}
=
\bigoplus_{i:\,u^i[s]=\sigma} u^i[y],
\end{align}
be the concatenation of all utterances within the utterance group, $\mathcal{G}$, spoken by the same speaker, where $\oplus$ denotes sequence concatenation.

In SD-CTC the loss is
\begin{align}
\mathcal{L}_{\text{SD-CTC}}
= - \sum_{\sigma \in \mathcal{S}} \log p\left(y_{\sigma} \mid \mathbf{x}\right),
\end{align}
which corresponds to modeling the speaker transcripts as conditionally independent given the input:
\begin{align}
p\left(\mathcal{G} \mid \mathbf{x}\right)
= \prod_{\sigma \in \mathcal{S}} p\left(y_{\sigma} \mid \mathbf{x}\right).
\end{align}

In contrast, the shuffle loss couples speakers through the shuffle product $\mathcal{Y}$:
\begin{align}
p\left(\mathcal{Y} \mid \mathbf{x}\right)
=
\sum_{a \in \beta^{-1}\left(\mathcal{Y}\right)} p\left(a \mid \mathbf{x}\right),
\end{align}
where the summation marginalizes over all alignments consistent with the shuffled token sequences. Despite the independence assumption in the loss, SD-CTC can still capture speaker interactions through the shared encoder, since speaker and token predictions are produced from the \emph{same} encoder representations via separate linear projections.

The speaker-specific blank used in SD-CTC is identical to the one used in our $N$-pass decoding method. Analogously, SD-CTC performs an $N$-pass \emph{training} procedure: speaker-specific logits are passed through separate supervision graphs, and the resulting path-sum losses are summed across speakers. Because non-target speaker tokens are merged into the blank symbol, the set of alignments considered by the shuffle product is a subset of those marginalized by SD-CTC. In this sense, SD-CTC can be viewed as a relaxation of the shuffle objective.

One significant difference between the shuffle loss and SD-CTC is that models using the shuffle product  do not require speaker labels. They can therefore more generally be used in tasks where an input is annotated only with a set of sequences but no speaker labels. For instance, one might have annotated a transcript as well as a sequence of sound events present in an audio recording, a common occurrence in web video corpora~\cite{li2023yodas}.

\subsection{Aligning Overlapped Speech}  
In this final section of our method, we show how the shuffle product can be used for single-pass forced alignment of speech to a set of transcripts, even when using models trained only on single-speaker speech. Force aligning overlapped speech is helpful for the preparation of multi-talker ASR corpora, among other tasks, but it can also be used to provide token alignments for methods such as t-SOT training \cite{kanda2022streaming}.
To our knowledge, there is little to no work on alignment of overlapped utterances, and current single-speaker alignment methods fail on overlapped speech, causing such examples to be discarded~\cite{cornell2020detecting, so2025performance}.

Recall that in the speaker attributed setting, $a^\prime$ corresponds to a length-$T$  sequence of either $\oslash$, or tuples of token and speaker, and $\mathcal{Y}^\prime$ is the speaker augmented shuffle product. Given an utterance group, $\mathcal{G}$, with transcripts, speaker labels, and possibly utterance start and end times, the goal of multi-talker alignment in overlapped speech can be framed using shuffles of the utterances as 
\begin{equation}
    {a^{\prime}}^* = \argmax_{a^\prime \in \beta^{-1}\left(\mathcal{Y}^\prime\right)} p\left(a^\prime | \mathbf{x}\right). \label{eq:alignment}
\end{equation}

Solving this problem can be accomplished by finding the Viterbi  path through, $T \circ G(\mathcal{Y^\prime})$. In the tropical semiring, the righthand side of Equation \ref{eq:forward_score} returns the probability of the best path, and the back-trace gives the minimizing path. Note that this is different from a normal alignment task where a single transcript is aligned to a corresponding audio recording. Here, we are aligning a whole set of transcripts to the recording.

\section{Metrics}
\label{sec:metrics}
We describe the metrics used to evaluate the speaker-attributed ASR and alignment methods introduced in the prior sections.

\subsection{Alignment Metrics}

\textbf{Boundary Error (BE)} -- We measure segmentation accuracy using the average boundary error across all tokens as in \cite{huang2024less}. Let $t_b^*[i,j]$ and $t_e^*[i,j]$ denote the reference start and end times of the $j$-th token in the $i$-th utterance, and $\hat{t}_b[i,j]$, $\hat{t}_e[i,j]$ the corresponding hypothesis times. The BE is computed as
\begin{equation}
\mbox{BE} = \frac{1}{N}\sum_{i=1}^N \frac{1}{2|u^i|}\sum_{j=0}^{|u^i|}
|\hat{t}_b[i,j]-t_b^*[i,j]| + |\hat{t}_e[i,j]-t_e^*[i,j]|.
\end{equation}

\textbf{Intersection over Union (IoU)} -- To account for varying segment durations we also report Intersection over Union (IoU), also known as the Jaccard index \cite{jaccard1901etude}, which measures the fraction of time two segments overlap relative to their union \cite{everingham2010pascal}. Similar metrics have been used for alignment under the name \emph{word temporal overlap ratio} \cite{deng2025speech, yang2022e2e}, though these normalize by the maximum segment duration rather than the union.

\textbf{Interleaving Distance} -- To measure whether tokens occur in the correct temporal order, we compute the Kendall-$\tau$ distance, defined as the number of adjacent swaps required to transform one ordering into another. We normalize by the number of reference tokens and denote this value by $\tau$.

\subsection{Transcription Metrics}

Multi-talker ASR is increasingly evaluated using metrics for speaker-attributed ASR such as the cpWER, ORC-WER, and MIMO-WER \cite{mimower}, as well as tools to compute them efficiently \cite{neumann23_chime}. Here we use the time-constrained minimum permutation word error rate (tcpWER)~\cite{von2025word} with a collar of $5$ seconds.

\section{Experiments}
\begin{table}[t]
\centering
\footnotesize
\caption{Statistics of evaluation datasets. Spk / s denotes the speaker density, or average number of active speakers per frame; overlap percentage denotes the fraction of time with more than one active speaker.}
\label{tab:datasets}
\begin{tabular}{lccccc}
\toprule
\multicolumn{6}{c}{Statistics of the Avg. Utterance Group by Datset} \\
\midrule
Dataset & \# Segs & \# Spks & Dur (s) & Spk / s & \% Ovlp \\
\midrule
2spk Ovlp
  & 8 & 2 & 39.1 & 1.4 & 42.9 \\
3spk Ovlp
  & 8 & 3& 35.4 & 1.7 & 55.0 \\
Libri2Mix
  & 2 & 2 & 8.4 & 1.6 & 64.2 \\
Libri3Mix
 & 3 & 3 & 9.0 & 2.2 & 72.2 \\
Synth
  & 2 & 2 & 9.2 & 1.5 & 52.7 \\
\bottomrule
\end{tabular}
\end{table}
We conducted experiments to examine (i) the effect of serialization schemes on speaker-attributed ASR; (ii) the relative performance of the proposed speaker attribution methods; (iii) models trained with the shuffle loss versus the SD-CTC loss; (iv) different methods for assigning speaker labels; and (v) the ability of these models to perform forced alignment to transcripts.

To study the behavior of our objective in a controlled setting, we used a simple experimental setup with simple models. Starting from pretrained speech encoders, we added only linear output projections to support the proposed objectives. To control the types of overlap seen during training and testing, we used synthetically overlapped LibriSpeech~\cite{panayotov2015librispeech} utterances. Table~\ref{tab:datasets} summarizes properties of the synthetic test sets, which contain a high degree of overlap since this is the focus of our work. We used the LibriMix~\cite{cosentino2020librimix} and Librispeech test sets for evaluation.
\subsection{Serialization}
\label{sec:serialization_experiments}
To what extent can a Transformer~\cite{vaswani2017attention} network, which theoretically allows arbitrary reordering of inputs, learn to serialize speech according to a deterministic serialization scheme? We compared deterministic serialization schemes with our proposed approach, which allows the model to select any serialization consistent with the temporal evolution of the acoustics under the partial-order supervision graph induced by different collars. We evaluated speaker SOT, t-SOT ($\kappa=0$), and progressively more flexible supervision graphs obtained by increasing $\kappa$. Decoding used the 1-pass algorithm from Section~\ref{sec:1-pass_decoding}.

All experiments used the pretrained WavLM~\cite{chen2022wavlm} Base model with the factored joint speaker attribution model described in Section~\ref{sec:speaker_attribution}. Encoder outputs were downsampled by 1D pooling (kernel=2, stride=2) to a 25\,Hz frame rate. Speakers were labeled by order of appearance (i.e., the first speaker to begin speaking is assigned label 1, the second label 2, and so on) and outputs were represented using a 5000-token BPE vocabulary trained on LibriSpeech.

Models were trained for 102k iterations with batches containing 280 seconds of speech. For the first 2000 iterations only the output projection layers were trained with a fixed learning rate of $5\times10^{-3}$. Subsequently, all parameters except the WavLM convolutional feature extractor were updated using Adam~\cite{kingma2014adam} ($\beta=(0.9,0.98)$, $\epsilon=10^{-8}$, weight decay $10^{-8}$) with a OneCycle scheduler \cite{smith2019super} (100k total steps, 8k warmup, maximum learning rate $10^{-4}$, cosine annealing, initial learning rate $5\times10^{-7}$, final learning rate $10^{-8}$).

Training data consisted of synthetic LibriSpeech mixtures constructed by sampling three speakers and splicing up to four of their utterances with partial overlap, constrained so that no more than three utterances were active simultaneously.
Each new utterance was mixed with a random SNR between -30 and 30 dB relative to the preceding one. Training began with single-speaker examples to improve pruning during lattice-based loss computation and gradually increased the minimum and maximum number of spliced utterances to two and four by 20k iterations. The maximum utterance group duration was 30 s.

Validation used similarly constructed, fully overlapped, two-speaker mixtures from LibriSpeech \texttt{dev-clean} and \texttt{dev-other}, but mixed at equal levels (0 dB). The number of checkpoints to average and blank penalty for decoding were tuned on two-speaker mixtures derived from \texttt{dev-clean} and \texttt{dev-other}, referred to as \textbf{Synth}. Training was performed on a single 32GB V100 GPU.

\subsection{Speaker Attribution}
\label{sec:speaker_attribution_experiments}
We then evaluate the performance of different speaker attribution models. In SD-CTC only the factored joint model was considered, so we compared it with the direct joint model described in Section~\ref{sec:speaker_attribution}. Because highly overlapped regions may require greater temporal resolution to separate speakers, we evaluated both attribution methods with and without the additional down-sampling described in Section~\ref{sec:serialization_experiments}; all other training settings were identical.
Due to space constraints, we summarize the main findings: both attribution methods performed comparably, but increasing the frame rate from 25\,Hz to 50\,Hz significantly improved performance on overlapped test sets while slightly degrading single-speaker performance, suggesting that finer temporal resolution helps resolve speaker interleaving.

\subsection{Shuffle versus Speaker Distinguishable CTC}
\label{sec:shuffle_v_sdctc_experiments}
As discussed in Section~\ref{sec:sdctc_relationship}, SD-CTC can be viewed as a relaxation of the shuffle loss. Although the two approaches differ in formulation, both can be decoded using the 1-pass method of Section~\ref{sec:1-pass_decoding}, so we compared models trained with these losses. For fair comparison, synthetic data generation was modified because our previous procedure occasionally produced overlapping segments from the \emph{same} speaker. 
While this acted as a form of regularization and improved shuffle loss performance, SD-CTC's per-speaker decomposition cannot handle same-speaker overlap.
Overlapping segments were therefore restricted to occur between \emph{different} speakers for fair comparison.

Mixtures were generated from up to 10 random segments drawn from up to four speakers. Each speaker was assigned a separate track, and segments were placed left-to-right between the end of that speaker's previous segment and the current end of the longest track, with at least 50\% overlap enforced. Track-level SNRs were sampled uniformly between $-15$ and
$15$\,dB relative to the first speaker. Mixtures were synthetically reverberated with probability $0.5$ and normalized to $-23$\,dB using pyloudnorm~\cite{steinmetz2021pyloudnorm} following~\cite{raj2024speaker}.

Experiments used a collar $\kappa=4$s, a 50\,Hz output frame rate, and a maximum utterance group duration of 50\,s. Because SD-CTC models speaker dependencies only through the shared network, we evaluated both WavLM Base and Large models to test whether additional capacity benefits SD-CTC relative to the shuffle loss. We also tested labeling speakers by speaking duration (total utterance length).

\subsection{Decoding}
\label{sec:decoding_experiments}
Using the best shuffle models from Section~\ref{sec:shuffle_v_sdctc_experiments}, we evaluated the effect of language model integration and oracle speaker count knowledge. Acoustic and blank weights were tuned on the \textbf{Synth} dev set (both set to 2.0) and applied uniformly to all test sets. We use the pruned (1e-07) 3-gram word-level language model released with LibriSpeech and a construct the lexicon from the BPE model.

\subsection{Alignment}
\label{sec:alignment_experiments}
To evaluate the ability of our models to align overlapped speech, we compared against a strong baseline and an oracle top-line. This alignment task differs from the standard setting because the audio must be aligned to multiple overlapping utterances. We consider two scenarios: (i) utterance boundaries are known and used to constrain the partial-order alignment FSA; (ii) utterance boundaries are unknown, i.e., for a $T$-second input, $\kappa \ge T/2$. Alignment is generally easier for shorter utterances since it requires less memory and fewer frames are available for token alignment. To demonstrate that the shuffle graph can handle longer inputs, we constructed two new test sets (2 and 3 speaker Overlaps) from the LibriSpeech validation data. Each utterance group contains eight mixed utterances from 2 or 3 speakers, averaging 35–40 s, compared to $\sim$10 s for Libri2Mix. We use alignments produced by the Montreal Forced Aligner (MFA), \cite{mfa}, as ground-truth. 

We assume that we always have some small amount of data to calibrate predictions to account for consistent bias in token start-times. We automatically calibrate the IoU and boundary errors by splitting the test set in two, using the first half for calibration. The Kendall-$\tau$ statistic is, however, computed over the whole test set.

\textbf{Baseline} – We test whether an ASR model trained on non-overlapped speech can be repurposed for overlapped speech using the shuffle graph. For this, we trained a speaker-attributed CTC model with the factored joint attribution model from Section~\ref{sec:speaker_attribution} on non-overlapped spliced LibriSpeech utterances, allowing the model to use both phonetic and speaker cues during alignment.

\textbf{Oracle} – Because the overlaps are synthetic, we can instead align each single-speaker track independently using a single-speaker ASR model and the CTC topology rather than the shuffle topology. This isolates errors caused by overlapping speech from those due to modeling choices such as BPE vocabulary size or the CTC topology.

\section{Results}
\label{sec:results}
\begin{table}[htb!]
\small
\centering
\caption{Test set results with tcpWER collar = 5s for WavLM Base at 25 Hz using the factored speaker attribution model
$p(\sigma_t \mid \mathbf{x}, \bar{\oslash}) \times p(\upsilon_t \mid \mathbf{x})$.
This table compares serialization strategies (Token Order). All values are tcpWER (\%).}
\setlength{\tabcolsep}{3pt}
\resizebox{\columnwidth}{!}{%
\begin{tabular}{@{}|c|c|ccc|cc|c|@{}}
\toprule
\multirow{2}{*}{\begin{tabular}[c]{@{}c@{}}\textbf{Token} \\ \textbf{Order}\end{tabular}} &
\multirow{2}{*}{\textbf{Synth}} &
\multicolumn{3}{c|}{\textbf{LibriMix Test}} &
\multicolumn{2}{c|}{\textbf{LibriSpeech}} &
\multirow{2}{*}{\textbf{Avg}} \\
\cmidrule(lr){3-5} \cmidrule(lr){6-7}
& & \textbf{3 Clean} & \textbf{2 Clean} & \textbf{2 Both} & \textbf{Clean} & \textbf{Other} & \\
\midrule
SOT            & 43.7 & 71.9 & 50.5 & 65.8 & 6.4 & 12.0 & 41.7 \\
\cline{1-8}
 $\kappa$ = 0s & 36.6 & 69.6 & 39.3 & 52.8 & 7.0 & 12.2 & 36.3 \\
 $\kappa$ = 1s & 20.7 & 61.4 & 24.1 & 41.5 & 6.0 & 10.6 & 27.4 \\
$\kappa$ = 2s & 19.2 & \textbf{60.5} & 22.3 & \textbf{38.7} & 5.8 & 10.5 & 26.2 \\
 $\kappa$ = 4s  & \textbf{18.5} & 60.6 & \textbf{21.1} & 39.7 & \textbf{5.7} & \textbf{10.4} & \textbf{26.0} \\
\bottomrule
\end{tabular}
}
\label{tab:serialization}
\end{table}
\begin{table*}[htb!]
\centering
\caption{Test set results with tcpWER (collar = 5s). All models use the factored speaker attribution model
$p(\sigma_t | \mathbf{x}, \bar{\oslash}) \times p(\upsilon_t | \mathbf{x})$ and a 50 Hz frame rate.
The table compares speaker ordering strategies and training losses. All values are WER (\%).}
\begin{tabular}{@{}c|c|c|c|ccc|cc|c@{}}
\toprule
\multicolumn{4}{c|}{\textbf{System}} & \multicolumn{6}{c}{} \\
\midrule
\multirow{2}{*}{\textbf{WavLM}}
& \multirow{2}{*}{\textbf{Loss}}
& \multirow{2}{*}{\textbf{Spk Order}}
& \multirow{2}{*}{\textbf{Synth}}
& \multicolumn{3}{c|}{\textbf{LibriMix Test}}
& \multicolumn{2}{c|}{\textbf{LibriSpeech Test}}
& \multirow{2}{*}{\textbf{Average}} \\
\cmidrule(lr){5-7} \cmidrule(lr){8-9}
& & & & \textbf{3 Clean} & \textbf{2 Clean} & \textbf{2 Both} & \textbf{Clean} & \textbf{Other} & \\
\midrule

\multirow{3}{*}{Large}
& shuffle & \multirow{2}{*}{start} & 12.7 & 60.9 & 17.6 & 30.2 & 5.4 & \textbf{7.9} & 22.5 \\
\cline{2-2}
& sd-ctc &  & 13.3 & 58.5 & 16.8 & 32.3 & \textbf{5.2} & 8.2 & 22.4 \\
\cline{2-3}
& shuffle & \multirow{2}{*}{length} & 11.8 & 44.3 & 8.1 & \textbf{25.3} & 5.3 & 8.2 & \textbf{17.2} \\
\cline{2-2}

& sd-ctc & & \textbf{11.1} & \textbf{43.8} & \textbf{7.8} & 28.2 & 5.3 & 8.2 & 17.4 \\

\midrule\midrule

\multirow{2}{*}{Base}
& shuffle & \multirow{2}{*}{start} & 15.1 & 53.3 & 10.6 & 37.5 & 5.3 & 9.9 & 22.0 \\
\cline{2-2}
& sd-ctc & & 15.4  & 53.4 & 10.6 & 40.4  & 5.6 & 10.4 & 22.6 \\
\bottomrule
\end{tabular}
\label{tab:base_v_large}
\end{table*}
\subsection{Serialization}
Results of the serialization experiments described in Section \ref{sec:serialization_experiments} are shown in Table \ref{tab:serialization}. We see that speaker-level serialized output training was difficult for the network to learn, resulting in the worst performance of all methods tried. Next, t-SOT, i.e., $\kappa=0$, was easier for the network to learn but resulted in the second worst performance. Presumably the heuristic sometimes enforces bad alignments which can degrade performance. 

We then use the partial order graph to relax the strict order of tokens and let the network learn to interleave overlapped utterances. Using a large enough collar ($\kappa =2s$, i.e., 2 seconds on either side), appeared to be sufficient for the model to significantly improve over the t-SOT single-target baseline $\kappa = 0s$. Interestingly, while performance on single speaker speech also improved using large $\kappa$, the change was less dramatic.

\subsection{SD-CTC vs. Shuffle Loss}
Table~\ref{tab:base_v_large} shows results comparing SD-CTC and the shuffle loss. An important observation, however, is unrelated to the loss functions themselves: the Base model performs substantially better on the overlapped test sets than the best Base model in Table~\ref{tab:serialization}. This improvement appears to come entirely from the modified data preprocessing, with loudness normalization likely playing the largest role. 
Interestingly, applying the same normalization to the test data has little effect, suggesting that the benefit comes from more consistent training dynamics rather than train-test mismatch.

Both loss functions perform comparably. The shuffle loss is slightly better with the Base model, while SD-CTC is marginally better with the Large model. SD-CTC is simpler to implement and somewhat more memory efficient, but it forces speakers to be treated independently during training. While the shuffle loss may be a more principled objective, we did not observe large performance differences between the two. It is possible that with highly overlapped real training data, where speaker discrimination is more difficult, the shuffle objective could provide greater benefit. The consistently stronger performance of shuffle models on the LibriMix 2 Both set hints at this, though the evidence remains weak.

A clearer result is that ordering speakers by total speaking duration (length) performs substantially better than ordering by time of first appearance (start), with improvements occurring entirely on overlapped test sets. However, this strategy is only applicable in offline recognition where the full utterance is available. As expected, the WavLM Large model also provides significant improvements over the Base model.

\begin{table}[htb!]
\centering
\footnotesize
\caption{Alignment results on synthetic LibriSpeech mixtures. Performance in terms of the IoU, BE in milliseconds, and the Kendall-$\tau$ distance, ($\tau$). Alignment error is measured against MFA word-level alignments on the corresponding single-speaker recordings.}
\setlength{\tabcolsep}{2.5pt}
\resizebox{\linewidth}{!}{%
\begin{tabular}{@{}c|c|c|ccc|ccc@{}}
\toprule
\multirow{2}{*}{\begin{tabular}{c}\textbf{Align} \\ \textbf{Graph}\end{tabular}} &
\multirow{2}{*}{\begin{tabular}{c}\textbf{Collar} \\ ($\kappa$)\end{tabular}} &
\multirow{2}{*}{\textbf{Model}} &
\multicolumn{3}{c|}{\textbf{2spk overlap}} &
\multicolumn{3}{c}{\textbf{3spk overlap}} \\
\cmidrule(lr){4-9}
& & &
\textbf{IoU $\uparrow$} &
\textbf{BE $\downarrow$} &
$\mathbf{\tau}$ \textbf{\% $\downarrow$} &
\textbf{IoU $\uparrow$} &
\textbf{BE $\downarrow$} &
$\mathbf{\tau}$ \textbf{\% $\downarrow$} \\
\midrule

$G\left(\mathcal{Y^\prime}\right)$ 
& 2s & \multirow{1}{*}{CTC} &  56.2 & 89 & 13.8 & 45.0 & 138 & 41.2 \\

\midrule\midrule

\multirow{4}{*}{$G\left(\mathcal{Y^\prime}\right)$} & 2s
& \multirow{2}{*}{Shuffle} & 61.8 & 69 & 6.3 & 58.2 & 81 & 16.9 \\

 & 32s
&  & 61.7 & 70 & 6.2 & - & - & - \\

\cline{2-9}

 & 2s & \multirow{2}{*}{SD-CTC} & 63.4 & 66 & 6.2 & 59.5 & 78 & 16.9 \\
& 32s &  & 63.2 & 66 & 6.2 & - & - & - \\

\midrule
\midrule
$G\left(u\left[y\right]\right)$ & N/A & CTC & 64.6 & 63 &  5.7 & 64.2 & 65 & 10.1 \\

\bottomrule
\end{tabular}
}
\label{tab:overlap_alignment_methods}
\vspace{-4mm}
\end{table}
\begin{table}[htb!]
\small
\centering
\caption{Test set word error rates (\%). All results are tcpWER (collar=5s).
}
\setlength{\tabcolsep}{3pt}
\resizebox{\columnwidth}{!}{%
\begin{tabular}{@{}l|c|ccc|cc|c@{}}
\toprule
\multirow{2}{*}{\textbf{Decoding}} 
  & \multirow{2}{*}{\textbf{Synth}} 
  & \multicolumn{3}{c|}{\textbf{LibriMix Test}} 
  & \multicolumn{2}{c|}{\textbf{LibriSpeech}} 
  & \multirow{2}{*}{\textbf{Avg}} \\
\cmidrule(lr){3-5} \cmidrule(lr){6-7}
  &        & \textbf{3 Clean} & \textbf{2 Clean} & \textbf{2 Both} 
           & \textbf{Clean} & \textbf{Other} 
           &                \\
\midrule
Greedy      & 11.8 & 44.3 & 8.1 & 25.3 & 5.3 & 8.2 & 17.2 \\
tgt spk HLG & 9.6  & 42.5 & 6.2 & 24.2 & 3.6 & 6.1 & 14.8 \\
\ \ + oracle \# spks & 9.6  & 42.5 & 6.2 & 24.2 & 3.6 & 6.1 & 14.8 \\
\midrule
UME \cite{shakeel2025unifying} & - & 15.9 & 6.4 & 19.6 & - & - & - \\
Yang MTASR \cite{RobustMTASR} & - & - & 5.1 & - & - & - & - \\
DiCoW\cite{Polok2024DiCoWDW} & - & 47.1 & 16.2 & 21.6 & - & - & - \\
SE-DiCoW\cite{polok2026se} & - & 29.3 & 3.4 & 8.7 & - & - & - \\
GEncSep \cite{GEncSep} & - & 13.1 & 6.6 & 15.0 & - & - & - \\
SOP \cite{shi2025serialized} & - & 16.5 & 3.6 & 9.2 & - & - & -\\
\bottomrule
\end{tabular}
}
\label{tab:simple_results}
\vspace{-7mm}
\end{table}
\subsection{Decoding}
We also evaluate the effect of the proposed multi-talker decoding algorithm. Incorporating a language model via target-speaker WFST decoding substantially improves performance across all test sets. We further experimented with limiting decoding to a known maximum number of speakers, ignoring outputs from later ones. This brought only marginal gains, suggesting the model is already fairly capable of speaker counting. However, the effect may be understated because training used only four speakers, so at most one additional speaker needs to be suppressed. We expect the benefit of oracle speaker count information to be larger when training with more speakers.

For comparison, we also report results from recent multi-talker ASR systems evaluated on overlapped speech. Most of these systems are not directly comparable: they use substantially larger models, more supervised data, and models based on Whisper or large-language models (LLMs) may have been trained on audio or text overlapping with the test sets~\cite{tseng2025evaluation}. The one exception may be GEncSep~\cite{GEncSep}, which also trains a dedicated model from scratch on top of features extracted from WavLM Large. The Shuffle model is among the best results on the clean LibriMix 2-speaker set, but significantly worse on the 3-speaker set. It is possible that the GEncSep neural architecture may be more amenable for overlapped speech recognition and its use in conjunction with the shuffle model should be explored in future work. 

\subsection{Alignment}
Table~\ref{tab:overlap_alignment_methods} shows the performance of the proposed methods for aligning overlapped speech. The last row corresponds to the \textbf{oracle} model, trained with CTC on clean single-speaker speech and aligning each track in the mixture independently. The resulting average boundary error of 63\,ms corresponds to roughly a quarter to half a syllable, assuming an average English speaking rate of about 5 syllables/s (200\,ms per syllable) \cite{baese2015speaking}. With greater overlap, the Kendall-$\tau$ distance increases slightly due to a higher likelihood of token collisions and swaps.

The top row shows the speaker-attributed CTC baseline trained on non-overlapped speech but decoded on overlapped mixtures using the shuffle graph. Alignment performance degrades across all metrics, with the average boundary error increasing by nearly 50\% and IoU dropping from 64.6 to 56.2. Nevertheless, a BE of 89\,ms is reasonably small ($\approx \frac{1}{2}$ syllable), indicating that the shuffle graph enables alignment of overlapped speech even with a model not trained on any.

The middle rows show models trained directly on overlapped speech using either the shuffle loss or SD-CTC. Their alignment metrics are nearly identical to those of the oracle model on single-speaker speech. When utterance boundaries are not used ($\kappa=32$s), alignment performance degrades negligibly despite the much larger shuffle graph. This could facilitate annotation of speaker boundaries for diarization corpora if speaker marked transcripts are available. However, for the 3-speaker overlap set the resulting search space was too large to fit in GPU memory, causing alignment to fail in that setting.

\section{Conclusion}
\label{sec:conclusion}
We developed a novel method for modeling multi-talker ASR using shuffles and partial order FSAs, which enables alignment via the Viterbi algorithm. We proposed alternative speaker attribution models, and speaker labeling schemes (order of appearance vs. order of duration). We also described and evaluated two methods to perform multi-talker decoding, and provided what we believe to be the first single-pass method for multi-talker alignment of speech. Results on LibriMix indicate that shuffle models enable transcription of overlapped speech as an interleaved sequence. Our method provides a fully end-to-end, principled, and light-weight method for multi-talker ASR that is close to matching alternative approaches on some of the LibriMix benchmarks.  
Beyond speech, the framework should be applicable to any sequence labeling task where multiple sources are observed through a shared channel, such as polyphonic music transcription or multiplexed communication signals.

\section{Generative AI Use Disclosure}
Generative AI tools have been used only to help revise and refine the manuscript.


\bibliographystyle{IEEEtran}
\bibliography{refs}

\end{document}